# Sequential acquisition of fluorescence signals with changing fluorophore concentrations. Multivariate Curve Resolution with time measurements assistance.

Gabriel Siano*[a,b], Sofía Mora[b], Agustina Schenone[b,c], Leonardo Giovanini[a,d]

(a) Instituto de Investigación en Señales, Sistemas e Inteligencia Computacional (sinc(i), UNL-CONICET), Argentina

(b) Facultad de Bioquímica y Ciencias Biológicas, Universidad Nacional del Litoral, Argentina

(c) Instituto de Desarrollo Tecnológico para la Industria Química (INTEC, UNL-CONICET), Argentina

(d) Facultad de Ingeniería y Ciencias Hídricas, Universidad Nacional del Litoral, Argentina

* Corresponding author: gsiano@sinc.unl.edu.ar (G. Siano), Tel: +54 0342 4575233 (117)

Postal address: Ciudad Universitaria UNL, Ruta Nacional N° 168, km 472.4, FICH, 4to Piso (3000) Santa Fe – Argentina

# Abstract:

Sequential registering of fluorescence signals in conventional Excitation-Emission Matrices (EEMs), followed by modeling based on multilinear properties of the data, requires stable fluorophore concentrations throughout the acquisition of each EEM. Rapid concentration changes, as seen in chromatography or certain kinetics, can disrupt the conventional bilinearity of EEMs. This deviation depends on the relative rates of concentration changes versus spectral scanning speeds.

Although entire scans can be slow, individual data points are acquired almost instantaneously, maintaining linear dependencies. Within specific time intervals, concentrations can be assumed constant. By using time-based localization, partial EEM data can be organized into partially filled cubes that allow for the correct modeling of third-order data. Additionally, Multivariate Curve Resolution requires reshaping operations.

Two third-order datasets were analyzed using a time-assisted MCR implementation, following a strategy similar to one reported for chromatographic data (LC-EEM) with Parallel Factor Analysis. The second dataset comes from Diclofenac reaction kinetics (Kin-EEM). The results suggest that time-assisted MCR resolves such data effectively. The solutions found were similar to those obtained when processing the same data as cubes. Predictions from calibration models based on MCR results were comparable to those obtained when deriving higher order models from the same data. High degrees of similarity were achieved between resolved and reference spectral profiles. Both chromatographic and kinetic profiles were accurate and physically meaningful. The proposed strategy highlights the potential of incorporating time-based localization to enhance the analysis of fluorescence data in dynamic systems.

# 1. Introduction

In the present study, Multivariate Curve Resolution (MCR) [1], assisted with time measurements (MCR(t)), was implemented to analyze fluorescence data previously resolved using PARAllel FACtor (PARAFAC) [2] analysis or derived models [3], particularly those accounting for the existence of missing data [4]. The first dataset, referred to as LC-EEM [5] (Liquid Chromatography-Excitation Emission Matrix), corresponds to the analysis of samples containing Pyridoxine, using liquid chromatography coupled with a spectrofluorometer for the sequential recording of Emission (EM) variables in emission spectra (spEM) at different Excitation (EX) variables. The second dataset, Kin-EEM [6,7] (Kinetics-EEM), corresponds to samples containing Diclofenac, with the production of a by-product through photodecomposition and the recording of spEM (with one out of every two spEM in reversed orientation) at different excitations for the study of kinetics directly in a cuvette. In both cases, the analytes were quantified in the presence of potential interferents, with overlaps in the emission, excitation, and chromatographic or kinetic modes. Although the methodologies were not identical, time measurements corresponding to the fluorescence readings were obtained in both cases. These time measurements, specific to each sample, can be incorporated into N-linear models such as PARAFAC, for example, through the temporal localization of information and the implementation of smoothness constraints in some model profiles [5], particularly in those where smoothness has a real physical meaning (chromatographic and kinetic profiles in this case).

Generally speaking, a collection of fluorescence data obtained at different EX wavelengths, through consecutive and sequential EM spectral scans [8–10] (without simultaneous recording of all variables), can be arranged into a two-way data array. However, this arrangement conventionally referred to as a Fluorescence EEM, does not necessarily imply the expected bilinearity that is almost automatically associated with EEMs in the literature. In other words, the mathematical object can be generated as a two-way array filled with fluorescence data, but such an array will not necessarily possess the mathematical properties expected of certain matrices containing fluorescence data, such as conventional EEMs, from which bilinearity is typically anticipated to extract information from the data. The property of bilinearity will exist only if, during the scanning of an entire EEM (i.e., from start to finish), there are no concentration variations in the fluorophores. This will primarily depend on the relative speeds of data acquisition and the rate of variation in fluorophore concentrations. If the concentrations vary rapidly, by the time a single unit of information for the model (e.g., a "matrix") is obtained, the last recorded responses will be linear with respect to the concentration at the time of their measurement, but not with the concentration at which the initial measurements were linear (and vice versa). As a result, the obtained matrix will not be bilinear, regardless of whether or not it is considered an EEM. Consequently, models such as MCR or PARAFAC should not be used.

An example of this can be observed in **Figure 1**. The data presented were processed as conventional EEMs and assuming conventional bilinearity for the LC-EEM case. Specifically, these results were obtained from an MCR model for type 1 non-quadrilinear third-order data [9,11] (assuming independence between Excitation and Elution Time modes). In this approach, each conventional EEM was unfolded into a vector (common mode), and each vector was associated with a point in the resolved chromatographic profiles of each sample (augmented mode: samples and chromatographic profiles).

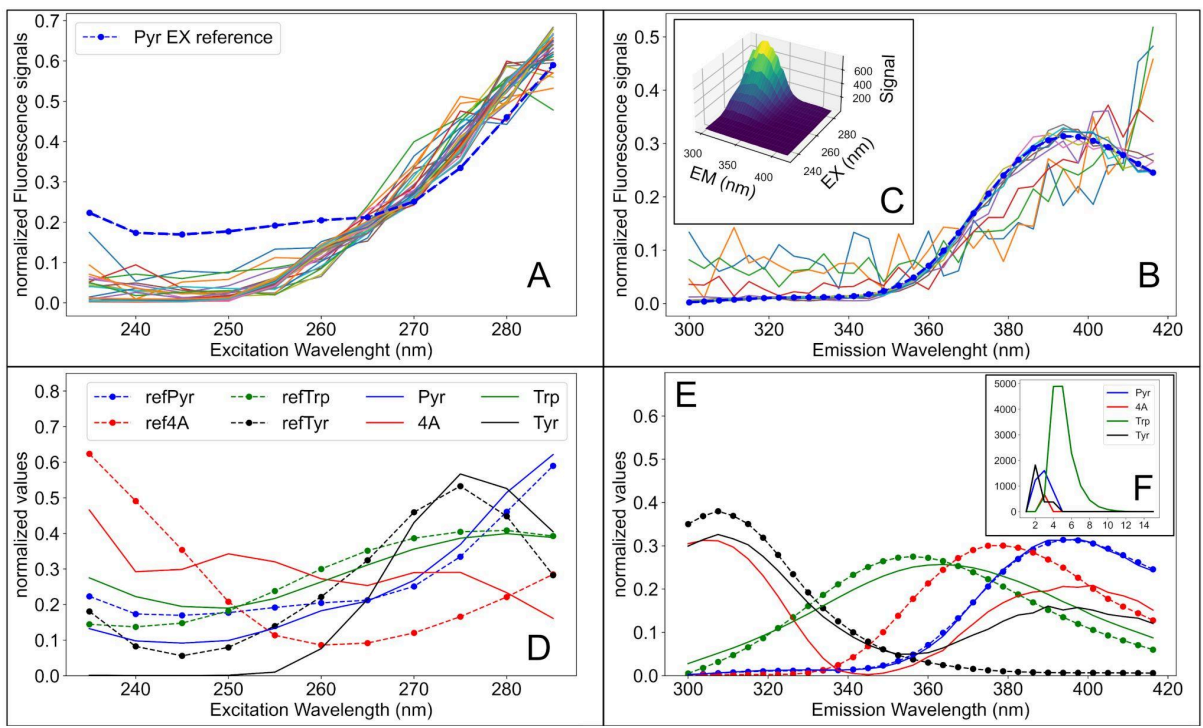

**Figure 1**: **Results of MCR models with the conventional interpretation of EEM for the LC-EEM dataset**: A) Normalized Excitation spectra (a color for each EM variable) obtained from a pure Pyr sample EEM, and the reference Excitation spectrum for Pyr. B) Normalized Emission spectra (a color for each EX variable) obtained from the same sample EEM as in A), and the reference Emission spectrum for Pyr. C) Sequential fluorescence scans initiated 30 seconds after sample injection, plotted as a conventional EEM (3 out of 15 total), for the sample in A) and B). D) and E) Resolved profiles by MCR without time assistance (solid lines) for a mixture of Pyr, 4A, Trp, and Tyr, showing Excitation (D) and Emission (E) profiles alongside their references (dashed lines). F) Chromatographic profiles resolved by MCR without time assistance (x-axis: EEM index, ordinal), for the same mixture of Pyr, 4A, Trp, and Tyr. Pyr: Pyridoxine; 4A: 4-aminophenol; Trp: Tryptophan; Tyr: Tyrosine; prefix "ref": reference.

In **Figure 1**, it can be observed that if the conventional EEM (the third of 15 total) in C is processed to obtain normalized Excitation (A) or Emission (B) spectra, respectively, these do not match, nor do they coincide with their references. Since these results correspond to a pure Pyridoxine sample, one would expect that, if the EEM were bilinear, the normalized EX and EM spectra would align with their references. It is clear that these EEMs were not even conventionally bilinear, not even for pure substances. This can be explained by observing F, where it is evident that concentrations varied while matrix 3 of 15 was being recorded. It is important to keep in mind that this type of scanning (the most common in general) requires time, and that the information modes can change simultaneously over time. This generates dependencies among modes, sometimes incompatible with multilinear models. In the typical case of MCR, the consequence of assuming bilinearity where it does not exist, and relying on this assumption to model the data, is the poor quality of the resolved profiles. The deficiency in the estimated chromatographic profiles can be summarized as a low number of points per profile with minimal physical significance. The spectra were not well calculated, with an average similarity criterion [12] of 0.945 for EX spectra and 0.867 for EM spectra. The model explained only 94.0% of the variance, with a Relative Error of Prediction (REP) of 29.9% and a mean recovery of 70.1% for the validation set. Further details can be found in Table S1.

Given that in these cases the matrices were not bilinear, it is important to understand that it is expected for a conventional MCR model to fail. The condition of being inappropriate arises from the relationship between the type of data and the type of model in their conventional forms, and thus cannot be improved through any combination of commonly used constraints. It is also worth noting that when these conventional MCR models were initialized with the known spEM and spEX for each substance, they became incorrect over the course of the iterations as the concentration profiles were adjusted. It is evident that, for this type of model, those reference spectra would not correspond to the best fit to the data.

Even with the aforementioned considerations, from a non-conventional perspective, even the trilinearity reported data persists to some extent. If, instead of viewing the EEMs as a unit of information, their individual points are considered, it can be concluded that during the recording of each single read/point (LC-EEM: 12.5 ms/read and Kin-EEM: 60 ms/read), it is reasonable to assume that the concentrations remained constant within each small time interval (i.e., c(t), concentration in time approximately constant). Logically, during each individual recording, both EX and EM variables also remained constant. Given that c(t), EX, and EM remained constant, it is possible to assume that each fluorescence reading was linearly dependent on these variables (excluding any additional phenomena). Similarly, instead of considering a single point, if a group of P consecutive points from the same scan is considered, the total times will be P·12.5 ms and P·60 ms for LC and Kin, respectively. Again, it is reasonable to assume that if the product of P and the time per reading results in short intervals, relative to concentration variations, then c(t) can be assumed to remain nearly constant, while EX and EM will have been determined by the instrument. In summary, the data will be trilinear either individually or even collectively within certain time scales. It should be noted that the times between reads within the same scan are much shorter than those between reads from different scans [5–7]. This is because changing scans requires the optical system to be reset and/or repositioned. These inter-scan times are also irregular (as verified for LC-EEM and Kin-EEM), due in part to handshaking protocols between the instrument and its control PC, which depend on the operating system and are entirely beyond the control of the operator. For this reason, it is advisable to measure or estimate the times of each fluorescence signal and incorporate them into the modeling. This allows for a coherent assumption of the time intervals over which the data can be considered multilinear. Without these time measurements, the modeling will implicitly assume equidistance between measurements, which may not always reflect the experimental reality.

Given all the above, since individually or in small groups of consecutive variables from the same scan, the data can be considered linearly dependent on the c(t), EX and EM variables, it can be useful to think in terms of pseudoEEMs or psEEMs rather than conventional EEMs. These psEEMs are matrices with the same dimensions as the original EEMs but containing only fragments of their information. Each fragment would be associated with a short time interval (to approximate constant c(t) ), while the rest of each psEEM would consist of missing data. Using these psEEMs, it is possible to construct 3-way arrays (cubes) for each EEM, partially filled with information, with known localization in the time domain, and where the data can reasonably be assumed to be multilinear. We refer to these incomplete cubes as psCubes. These psCubes, after reshaping and augmentation, will form the pseudoAugmentedMatrices for MCR. An example of psCubes for the LC-EEM system is illustrated in **Figure 2**.

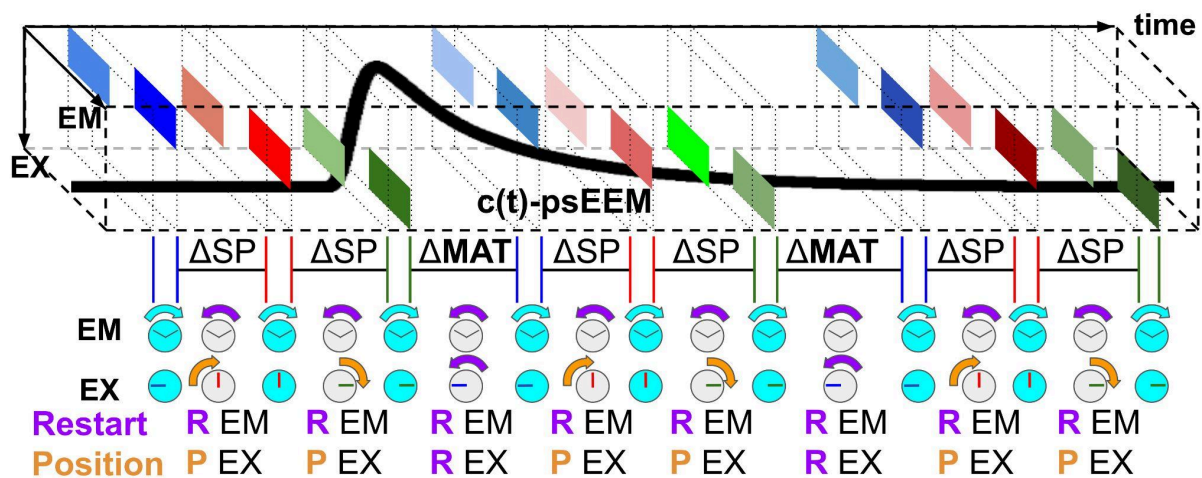

**Figure 2**: **Scheme of monitoring a chromatographic run using psCubes with EM scans.** Top**:** Three consecutive EEMs viewed as three consecutive psCubes, each composed of six psEEMs. Each psEEM, outlined with dashed lines, contains only half of an experimental spEM (colored rectangles representing the initial or final half of a scan) for a single EX variable (shades of blue, red, or green). For a single EX, different shades of the same color are used to represent signals with varying intensities, depending on concentration changes and/or whether it corresponds to the initial or final part of an spEM. The remaining data are considered missing data. Behind these, a hypothetical chromatographic profile is depicted with a thick black line. Bottom**:** Time differences between spectra (ΔSP) and between matrices (ΔMAT), along with schematics of the restarting and positioning of hypothetical EX and EM monochromators. Experimentally, neither ΔSP nor ΔMAT were regular.

**Figure 2** illustrates the recording of 18 psEEMs, each containing 3 EX variables (shades of blue, red, and green). The number of EM variables is undefined, but in the example, each psEEM retains half of an spEM. With a conventional interpretation, the first 6 psEEMs would correspond to the first EEM, psEEMs 7 to 12 to the second EEM, and psEEMs 13 to 18 to the third EEM. These 3 EEMs could be presented sequentially in time, but without being able to clearly define the exact times at which the readings occurred or the concentrations associated with those times. With this approach, the description of the chromatographic profiles would consist of only 3 points, which would be insufficient to represent rapid concentration variations (such as chromatographic peaks). In contrast, each psEEM in a model corresponds to a brief and specific time interval during which it can be reasonably assumed that the concentrations of the fluorophores remained constant (i.e., a form of temporary steady states). This approach will result in more realistic descriptions of the c(t) profiles, with more points, reduced contribution to calculation error for the EX and EM modes, and the possibility of leveraging multilinearity, depending on the model. The figure also shows time measurements and their derived values: ΔSP (time difference between different spEMs) and ΔMAT (time difference between the end of one EEM and the start of the next). It is worth noting that, in the schematic, ΔSP and ΔMAT appear to have constant values, but as mentioned earlier, they are irregular [5]. This irregularity is related to the restarting and positioning protocols for EM and EX, illustrated in the lower part of the figure, as well as to handshaking protocols between the instrument and its PC. It should be noted that the concatenation of psCubes over time can also be interpreted as the psCube of each sample (or its "augmented" psCube).

It can be deduced that for 2 or more conventional EEMs per sample, the number of derived psEEMs contributing to each point in a c(t) profile is equal to nEEM·nEX·nEM/nVarCLFC. This corresponds to the original number of conventional EEMs (nEEM) multiplied by the number of EX (nEX) and EM (nEM) variables, and divided by the number of variables in an spEM for which the concentration is assumed constant during its recording (nVarCLFC, number of Variables with Constant Local Fluorophore Concentrations). In the example, nVarCLFC would be nEM/2, and for scans at constant speed, it would correspond to the time required to record half an spEM.

Given that the trilinearity of these data, as well as the quadrilinearity of arrays composed of these data, has already been discussed in PARAFAC models or their derivatives [5–7], the present study focuses on data processed with MCR in its bilinear version only, to evaluate the effect of incorporating time measurements during modeling (the reader can find further details on models with third- or fourth-order relationships in the cited references). In this regard, it is also important to emphasize that there are multiple ways to apply multilinearity constraints [13–17]. However, this multiplicity suggests that there is no consensus on the absolute generalizability of any of these implementations, and evaluating them falls outside the scope of the present study.

Finally, since each sample contains third-order data (c(t), EX, EM), different unfolding strategies can be employed to obtain matrices for MCR (beyond the subsequent stacking of these matrices into a super-augmented matrix). In the common mode, it is common practice to retain the information mode with the lowest likelihood of variation. For these datasets, it would be EM. On the other hand, c(t) will undoubtedly be in the augmented mode, given the irregularities in timing and potential artifacts in the c(t) modes of each sample. Regarding the EX mode, it has been concatenated with EM in the common mode, as a time-assisted modification of MCR has been employed, and with this no dependencies between these modes were expected [9].

## 2. Material and Methods

2.1 Datasets

Two datasets were used, both consisting of spEMs taken at different EX wavelengths, with varying fluorophore concentrations. In both cases, the time of each fluorescence scan was recorded by connecting the respective spectrofluorometer to an Arduino board [18] with lab-written firmware. These time measurements were subsequently used in each iteration of MCR, specifically to smooth the c(t) mode of each component using the time measurements specific to each sample, as previously described with PARAFAC [5–7].

2.1.1 LC-EEM

The instrumental setup consisted of an Agilent 1260 UHPLC connected to a flow cell within a Cary Eclipse Fluorescence Spectrometer. Fluorescence measurements and their corresponding times were recorded for 21 samples. The calibration samples consisted of pure Pyridoxine (Pyr, vitamin B6) at different concentrations, while the validation samples also included 4-aminophenol (4A), Tryptophan (Trp), and Tyrosine (Tyr). For each sample, 15 conventional EEMs were recorded, spanning 11 EX variables (235 to 285 nm, in 5 nm increments) and 32 EM variables (300 to 420 nm, in 3.75 nm increments). Further details can be found in a previously published work [5].

2.1.2 Kin-EEM

A Perkin Elmer LS55 instrument connected to a PC running Python code developed in-laboratory for instrument control was used. A total of 30 samples were processed. The calibrated analyte was Diclofenac (DCF), which was photodegraded inside a quartz cell within the instrument. DCF was also quantified in other samples containing Trp and/or Pyr, both potential interferents. For each sample, 5 conventional EEMs were recorded consecutively, spanning 8 EX variables (275 to 292.5 nm, in 2.5 nm increments) and 64 EM variables (337 to 400 nm, in 1 nm increments).

As a result of studying certain variables [6,7], the pH of all samples was fixed at 3 using a phosphate buffer (0.9 M sodium phosphate monobasic - 0.1 M phosphoric acid), and the cell used for sample measurements was thermostatted at 20 ºC with a recirculating water bath. Additionally, based on bibliographic evidence suggesting that certain gases may actively participate in photodegradation kinetics [19] (and on experiments not reported here), it was decided to cover the measurement cell to prevent gas exchange. The values of the mentioned parameters, as well as other instrumental parameters (PMT detector voltage, EX and EM wavelength ranges, scan speeds, among others), were not strictly the result of an optimization process. Instead, these parameters, along with the sample concentrations, were selected to establish a reproducible experimental context, ensuring useful signals without associated issues (e.g., inner filter effect).

Two points are noteworthy: a) for the kinetics study, the irradiation source required to photodegrade DCF was the same excitation source of the spectrofluorometer. The degradation did not occur through additional irradiations but rather during the acquisition of fluorescence signals, b) one out of every two emission scans was obtained in reversed orientation (which must later be accounted for during the preprocessing of fluorescence and time data). The above was made possible because the instrument used is easier to program and control, although it is slower than the system used for LC-EEM. Regarding its speed and operation, Figure S1 shows characteristic time measurements and some derived statistical indicators, while Figure S2 provides a schematic representation of the orientation of each scan. This approach significantly reduced the time between spectra, from approximately 4.2 seconds to nearly 2 seconds, resulting in a total time saving of about 1.5 minutes per sample. Further details can be found in previously published works [6,7].

2.2 From EEMs to MCR super-augmented matrices

The first step in processing the experimental EEMs is to determine an appropriate value for nVarCLFC. That is, to identify the number of continuous variables within a spectral scan (spEM in this case) for which it can be assumed that fluorophore concentrations did not vary significantly. This can be achieved by implementing models with different numbers of grouped variables, from larger to smaller groups (an entire spectrum, half a spectrum, a quarter, etc.), and evaluating whether, beyond a certain reduction, no further improvements are observed in the models, but rather overfitting to the data [5]. It is important to note that the smaller the number of grouped variables, the greater the number of points in a time profile, but also the higher the computational resources required. Additionally, it is worth noting that the models can be refined gradually. That is, the results (concentration profiles and spectral profiles) from a model with many grouped variables can be adapted to initialize models with fewer grouped variables. From this procedure, which does not strictly represent an optimization process (as this falls outside the scope of this work and related previous studies), it was determined that the values of nVarCLFC that proved useful were 32 out of 32 (1 spEM) for LC-EEM [5] and 16 out of 64 emission variables (1/4 of spEM) for the Kin-EEM system [6,7]. Then, according to nEEM·nEX·nEM/nVarCLFC, each sample from the LC-EEM system was represented by an augmented psCube (see the schematic in **Figure 2**) consisting of 165 points (15·11·32/32) in non-equally spaced time intervals, 11 EX variables, and 32 EM variables. Similarly, the augmented psCubes derived from each sample of the Kin-EEM system had dimensions of 160 points (5·8·64/16) in time, 8 EX variables, and 64 EM variables. The schematic of an augmented psCube for the Kin-EEM system, along with specific details on scan orientations and waiting times, can be found in Figure S3.

To be processed in MCR, the augmented psCubes must undergo a refolding process into a two-way array. **Figure 3** illustrates a schematic of the basic structure used per sample in this study, for hypothetical data with 3 EX variables, 4 EM variables, and 3 EEMs.

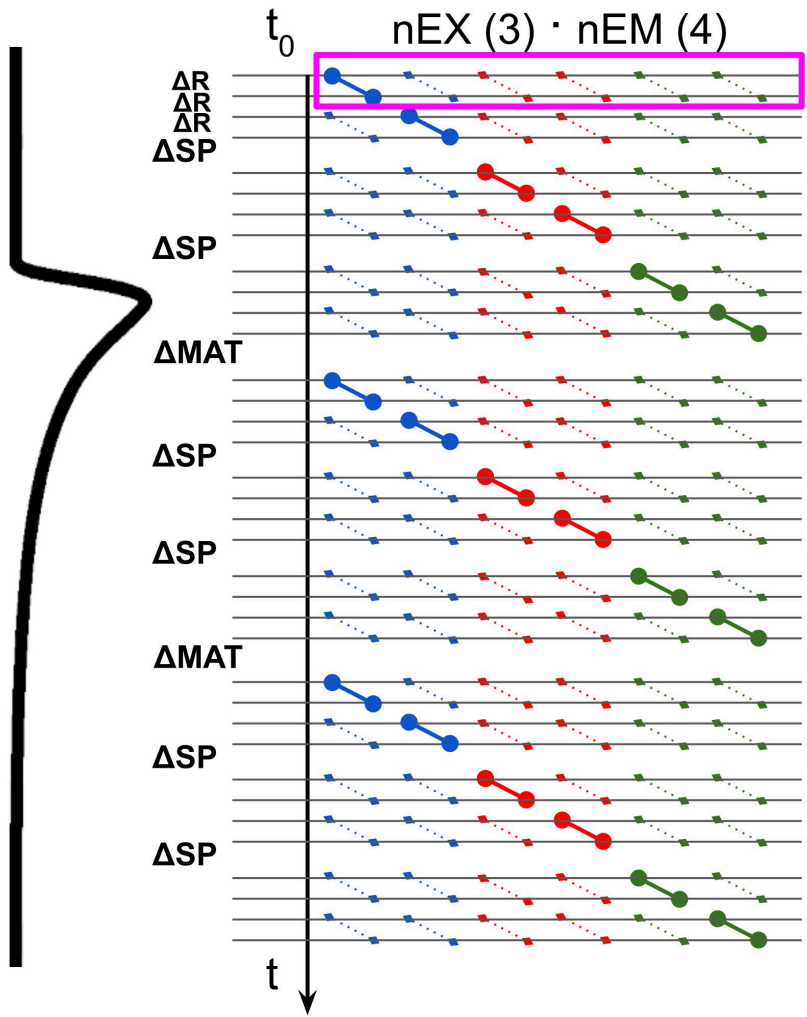

**Figure 3**: **Schematic of the data rearrangement from 3 conventional EEMs into two-way arrays for MCR, used to monitor the hypothetical chromatographic profile shown on the left.** Circles: experimental data; Diamonds: missing data, later imputed; t: time; $t_0$: start of the readings; nEX (3) and nEM (4): number of EX and EM variables. Blue, red, and green colors represent 3 EX variables. Magenta rectangle: the first psEEM of a cube, once unfolded (Δtime values are not to scale, ΔR << ΔSP < ΔMAT).

In **Figure 3**, a hypothetical chromatographic profile is shown on the left, which could just as well represent a kinetic profile in the case of Kin-EEM. The monitoring of these profiles would occur through consecutive EM scans at 3 EX wavelengths (blue, red, and green, in that temporal order), resulting in 3 conventional EEMs per sample. The magenta rectangle highlights what would be the first psEEM unfolded into a vector. It assumes that between the first and second readings (or between the third and fourth), a time interval elapses during which concentrations are considered constant. Only these experimental readings would constitute the first psEEM, while the rest of the data would be treated as missing data. Similarly, as time progresses (downward), each psEEM unfolded into a vector will contain a small fraction of experimental information, corresponding to one of the 3 EX variables and either the first or second half of an spEM. It is important to note that the experimental data are not aligned horizontally. If they were, they would correspond to data from an instrument capable of performing all measurements simultaneously (within a ΔR) rather than sequentially. The diagonal arrangement of the experimental data indicates that they are sequential combinations of EX and EM variables that depend on time. It is useful to compare the concentration variations in the profile on the left between two time points. If these points represent the start and end of either the first or second half of almost any spEM, the variations are generally slight. However, if they represent the start and end of an spEX, the concentration variations are significantly larger. This clearly illustrates the greater time dependence of spEX.

It is important to note that the proposed model requires complete horizontal vectors. This implies that, for the model, the recording of each psEEM is assumed to occur simultaneously within a ΔR. Over such a short time interval, it is assumed that no concentration changes occur and that neither EX nor EM depend on time.

At this point, it is useful to distinguish between experimental and imputed data. The latter are necessary to meet certain model requirements, both at the start and during the iterations. In this study, Expectation Maximization [4] was applied at each iteration, generating the missing data based on the partial results of the bilinear model from each iteration, similar to how this is implemented in the N-way Toolbox [20] for PARAFAC models of any order. Whether this or another imputation method is the best is not under discussion here. The approach simply seeks to approximate what would be expected if the data were complete. Nonetheless, it is important to clarify that the minimization of errors in the fitting function during the iterations only considers the experimental data and never the imputed ones.

Finally, it is worth highlighting that the schematic in **Figure 3** represents a single sample and that the ΔSP and ΔMAT values were always irregular. Therefore, a conventional MCR model would also need to be superaugmented, involving a stacking of samples in the augmented mode that accounts for both intra- and inter-sample time differences.

2.3 MCR Models details

2.3.1 Previous results

To obtain initial estimates for both datasets, the EX and EM spectra resolved with 4-way PARAFAC models assisted with time measurements were used. From previous studies [5–7], it was known that these data were not strictly quadrilinear, making it reasonable to expect that bilinear models could improve upon these initial estimates. That is, the spectral profiles of the 4-way model correspond to a single c(t) profile per component in all samples, whereas time-assisted MCR allows each sample to obtain its own c(t) profiles based on its specific time measurements. The initial estimates were also used to impute the missing data before the first iteration of each MCR model. Subsequently, through Expectation Maximization, the missing data were imputed with the bilinear model at each iteration. From the same previous studies, smoothing parameters [21] were reused for the respective c(t) modes, without assuming that the measurements were equidistantly spaced in time, but instead utilizing the specific time points of each sample to influence the smoothers in the MCR models. As has been previously stated [22], in this type of processing, where certain results are smoothed based on predictor measurements, the predictor is the time at which a measurement is taken. Except for the c(t) modes, no other type of smoothing was applied to the data or their profiles. A detailed investigation of the smoothing parameters falls outside the scope of this study.

2.3.2 Model constraints

For both datasets and in all modes, non-negativity was applied. For the augmented mode with c(t) profiles, unimodality and time-based smoothing were employed. For the calibration samples, selectivity (absence of non-calibrated components) was used. The base code was from the MVC3 Toolbox [23], with additional lab-written code developed to implement Expectation Maximization with the bilinear model during each iteration, as well as to smooth the concentration profiles based on the time measurements of each sample. Further details on the MCR models can be found in the literature [1].

# 3. Results and Discussion

3.1 Time-assisted MCR for LC-psEEM

**Figure 4** shows the results of MCR assisted with time measurements:

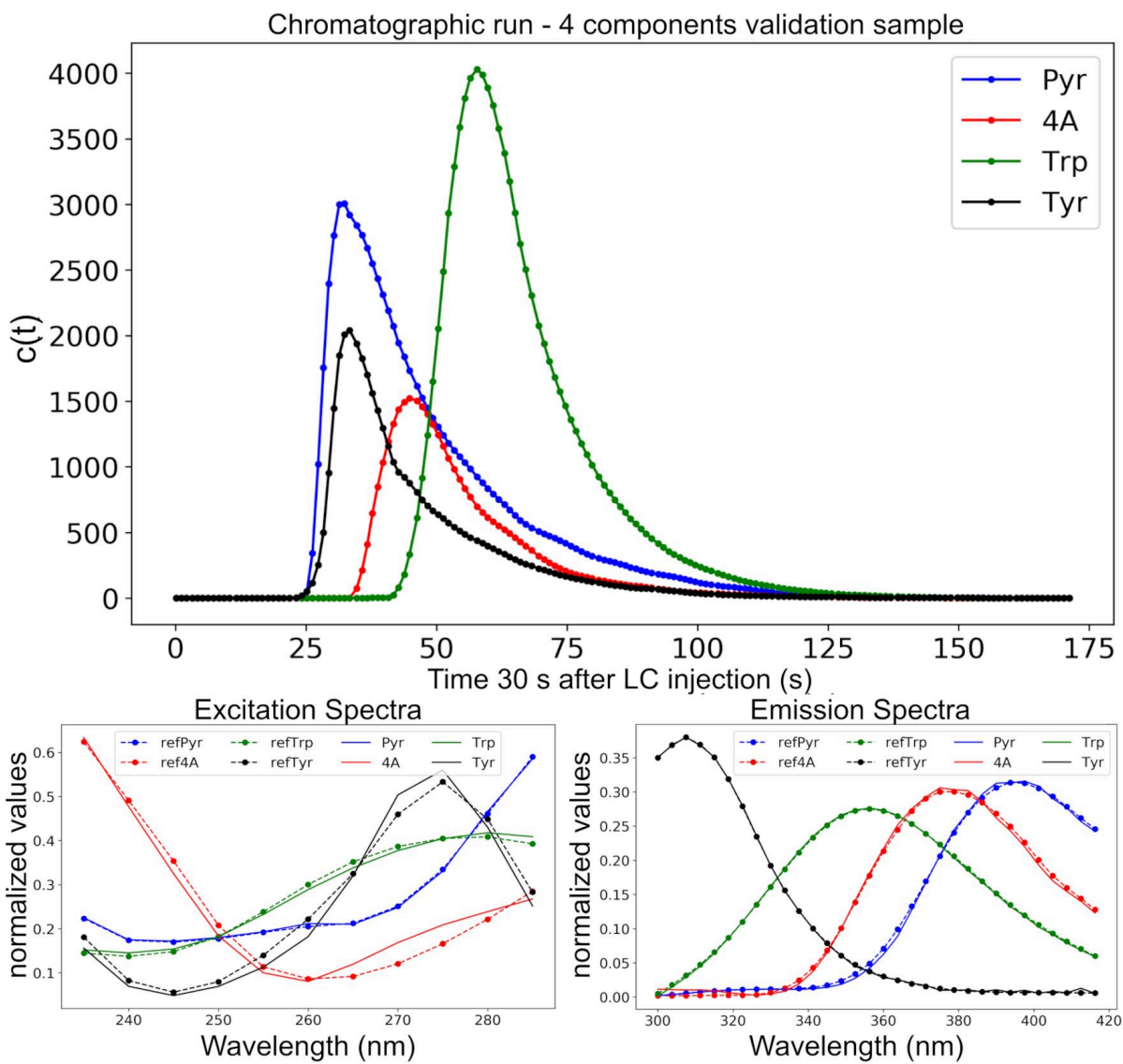

**Figure 4**: **LC-EEM time-assisted MCR model results for a validation sample with 4 components**. Top: chromatographic profiles resolution, bottom: spectral profiles resolutions (normalized values). Pyr: Pyridoxine; 4A: 4-aminophenol; Trp: Tryptophan; Tyr: Tyrosine; the prefix "ref" means reference.

As shown in **Figure 4**, the resolved profiles were really good. The explained variance was 99.6% (considering only the experimental data, not the imputed ones). The lowest similarity coefficients were 0.9963 for EX and 0.9995 for EM, both for 4A. The REP for the validation set was 1.76%, and the average recovery was 101.76%. Although with slight imperfections, the chromatographic profiles were obtained with significantly greater physical meaning, thanks to the time-based smoothing. These results were achieved with a missing data percentage of 90.909%. The resolution of the model with such a high percentage of missing data was partly made possible by the strong condition imposed by a structural relationship among the data, which was also used to impute the missing data. In this case, that relationship was bilinearity, although this alone is not sufficient. Another critical contribution was the use of time measurements combined with appropriate smoothing parameters. This approach enables the temporal localization of the data and the smoothing of c(t) profiles, which can be expected to exhibit genuine smoothness over time in terms of physical reality. The measured time can be quantitatively related to a physical phenomenon. Variations in c(t) are limited and will always exhibit a degree of smoothness. As has been said, this type of smoothing procedure can aid in estimating missing elements, particularly in longitudinal datasets where measurements occur within the same time span but vary across time points for different variables and/or occasions [22]. For the spectral modes, there is, a priori, no indication of smoothness related to time. This is one reason not to smooth the data beforehand, as there is no evident criterion for doing so.

In previous experiments [5], it was shown that for models of order higher than 2, while the implementation of psCubes with Expectation Maximization improves results compared to using EEMs directly to construct 3-way arrays for each sample, these results are not realistic due to artifacts in the c(t) profiles. However, if time measurements are incorporated to localize each psEEM within each psCube and these measurements are also used to smooth the c(t) profiles during each iteration, it is possible to fine-tune the smoothing parameters and obtain realistic solutions. Similarly, in this study, bilinear models derived solely from vectors of each psEEM were not sufficiently accurate (not shown), but incorporating time measurements and time-based smoothing yielded appropriate results. Additional details regarding the LC-EEM dataset can be found in **Table S2**.

3.2 Time-assisted MCR for Kin-psEEM

The c(t) profiles calculated by time-assisted MCR can be seen in **Figure 5**:

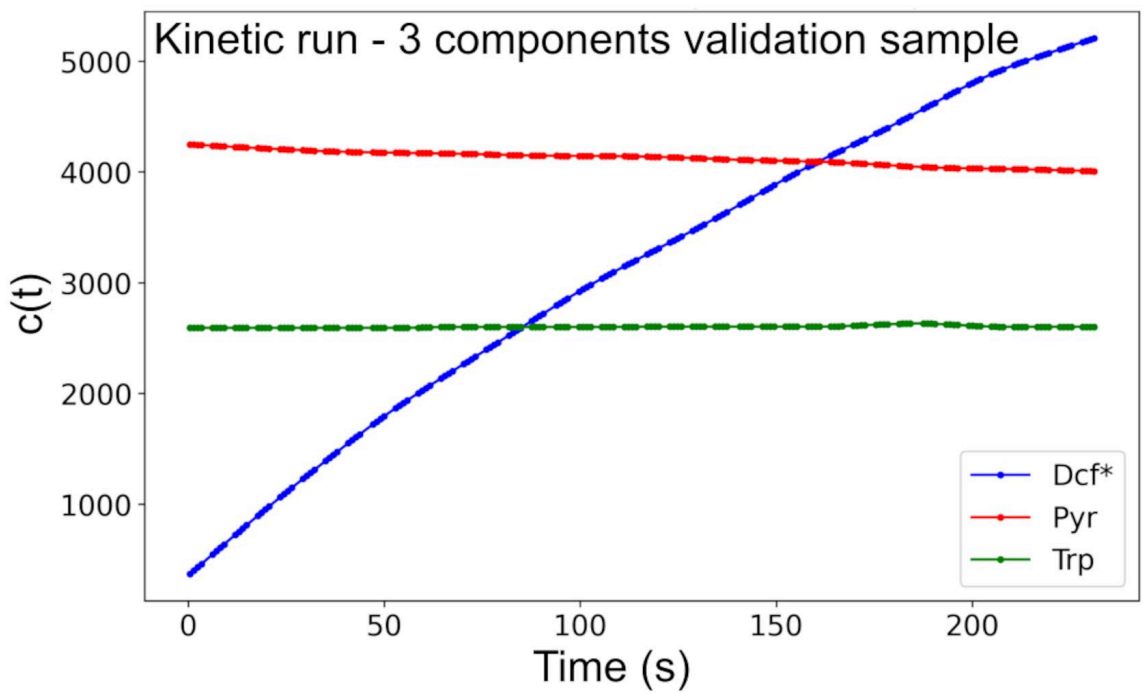

**Figure 5**: **Kin-EEM time-assisted MCR model concentration profiles for a validation sample with 3 components**. Dcf*: DCF by-product; Pyr: Pyridoxine; Trp: Tryptophan.

As seen in Figure 5, the case of the kinetic profile related to DCF, which was also the only detectable component in the pure DCF samples, consistently exhibited increasing kinetics over time in all samples. This, along with similar results obtained using time-assisted PARAFAC and others obtained via HPLC [6,7], suggests that at least one fluorescent by-product originates from the degradation of DCF, which itself may not be fluorescent under the experimental conditions or may be fluorescent at undetectable levels. Research has explored photodegradation pathways and by-products, though the process remains not fully understood [24]. The resulting spectral profiles exhibited high similarity coefficients (>0.99) with their references. The explained variance (considering only the experimental data) for the model was 99.8%, with a REP of 3.67% and a mean recovery of 96.33% for the validation set. The percentage of missing data was 96.875%, even higher than for LC-EEM. From a technical perspective, the data were acquired with one out of every two emission scans in reversed orientation. This reduced the restarting and repositioning times for the monochromators, decreasing ΔSP from about 4.2 seconds (non-reversed, conventional common-orientation scans) to approximately 1.9 seconds. This adjustment reduced the analysis time per sample by approximately 1.5 minutes, resulting in a total of around 3.85 minutes of analysis per sample.

In these experiments, the instrument excitation lamp itself caused the degradation of DCF while fluorescence readings were being acquired. This highlights the need to work within small time windows, ensuring minimal variations in concentrations. Additionally, considering that even the light required for excitation can lead to degradation, it is worth questioning whether it is always feasible to acquire an entire EEM that is conventionally bilinear. One could think of such an EEM as a special case of a psCube, one in which (in addition to the absence of saturations, quenching or other interactions, etc.): a) all the "fragments" necessary to reconstruct the data into a 2-way array are available, and b) all the "fragments" were recorded with constant fluorophore concentrations, even if this occurred at different times. This EEM can be considered a projection of a psCube onto the EX-EM plane, collapsing the time dimension under the assumption that concentrations remained constant. Considering that even the internal EX lamp can photodegrade substances, the assumption of constant concentrations, often just reported for designed samples, should ideally emerge as the result of an analysis using psCubes (or matrices derived for MCR) that explicitly account for time, including its irregularities. The general conclusion is that the most appropriate approach is to organize this type of longitudinal data into a structure compatible with its inherent properties (multilinearity in these cases), including the time at which the recordings occurred as a foundational element of that structure. All the profiles calculated by MCR assisted with time measurements can be seen in **Figure S4**, and further information can be found in **Table S3** and **Table S4**.

3.3 Time-assisted MCR generalities

Some considerations and conclusions are compatible with both datasets.

3.3.1 On the use of time measurements

It is important to note that time measurements must actively participate in the calculations to influence the results, rather than being used merely at the end to graphically present the c(t) profiles. For this reason, they must be used in each iteration as part of a model constraint to smooth the c(t) profiles, while accounting for the irregularities and temporal characteristics specific to each sample.

It is also worth highlighting the instrumental handling combined with chemometric techniques to achieve valuable analytical time savings. In Kin-EEM, time was saved by scanning one out of every two scans in reversed orientation. A more general saving comes from designing the models by selecting the appropriate time scale (LC-EEM and Kin-EEM). These types of details enable the extraction of information from rapid processes using hardware conventionally available in an analytical laboratory, without intentionally slowing down any system variable, which would otherwise compromise analytical efficiency.

3.3.2 On the unfolding of psEEM into vectors

To implement MCR with these data, it was necessary to unfold them in one of multiple possible ways to obtain 2-way arrays. The success of the resolution using the chosen unfolding strategy, as opposed to other possible methods that assume dependencies between modes, suggests that such interactions do not actually occur in the data. The unfolding of each psEEM into a vector implies that the resulting EX and EM modes are independent of the c(t) mode. As previously discussed, this holds true over short time intervals where c(t) remains nearly constant. It is also important to note that the results were achieved by leveraging only the bilinear model, despite knowing that the data are trilinear (at least partially).

Therefore, for this type of data, an interpretation based on a structure of incomplete data localized in time (psCubes or their derivatives via reshaping or unfolding), appears to be more suitable than an interpretation based on a conventional and forced structure of complete data, without appropriate time localization.

In addition, the structure of psEEMs unfolded into vectors would be the most appropriate given the nature of the data. Suppose there is an instrument capable of sequentially capturing multiple EEMs, with all their EX and EM variables recorded simultaneously (within a common and short integration time). During the acquisition of each EEM, concentrations would remain constant within certain limits, and thus each EEM would be conventionally bilinear. Therefore, they could be unfolded into vectors and stacked to form matrices, where each row would correspond to a specific time point. This would be equivalent to **Figure 3**, but without imputed data and with each EEM complete, containing all its experimental data.

3.3.3 On linear dependencies and the use of possible constraints

In this study, the aim was to evaluate the bilinear model assisted with time measurements, even though it was always understood that the data are trilinear (at least partially). While the results of the bilinear model appear sufficiently good with the proposed modifications, alternative results could potentially be obtained by incorporating constraints related to multilinearity. However, it is important to distinguish between its application relative to the common mode or the augmented mode. For the common mode, consider a sample of a pure substance and the recording of a complete EEM without concentration variations over time. An EEM of this type, when unfolded, would be equivalent to the Kronecker product of its spEX and spEM. That is, it would be a sequence of spEMs, placed side by side, with each scaled sequentially by an EX coefficient. This implies that a relationship indeed exists between spEX and spEM, and therefore, it could be meaningful to incorporate this relationship as a constraint in the model. As shown in this study, the bilinear model alone has proven sufficient for these data, and no additional constraints were required to achieve satisfactory results. On the augmented mode side, a trilinearity constraint could logically also be applied to each sample. However, applying a quadrilinearity constraint would imply that all the c(t) profiles of a given component across different samples must have the same shape and be aligned within the same chromatographic time or kinetic interval. Although there are algorithms capable of modeling changes in shape and/or time (depending on the implementation of the constraint), the use of time-assisted modeling allows for respecting the natural characteristics of the measurements without forcing the data to fit predetermined profiles. While it is understood that this could be useful in certain cases, it does not seem particularly appropriate for our study. For the data described here, we know with certainty that shifts and distortions exist in the c(t) modes. Moreover, after modeling, we confirmed that the data are not strictly quadrilinear [5], precisely due to small temporal variations in the profiles for each sample. It must be understood that when the occurrence times of all fluorescence readings have been measured, the necessary information is available to process the c(t) profiles of each sample according to their specific shape and position. Therefore, it is not necessary to seek general characteristics (shapes and/or times) across samples in the concentration profiles. Instead, the specific recording sequence of each sample and its differences relative to others are respected, with shared information confined to the common mode. Additionally, the algorithms dedicated to such tasks typically assume regularity and equidistance between measurements. Moreover, conventional implementations of PARAFAC and MCR also make the same assumption (as in the MCR model shown in Figure 1). Therefore, for data like those presented here, even if they are multilinear, it will be very challenging for a multilinearity constraint to effectively leverage the multilinearity of the data if the structure distorts how this property would normally be expressed.

3.3.4 On estimating initial profiles and imputations in MCR

It is important to highlight that, to achieve good resolutions with the bilinear MCR models, it was necessary to modify the algorithms to accommodate time measurements, as well as handle and impute missing data. Furthermore, both the initial imputations and the initial profile approximations were based on prior results from PARAFAC. Without these prior results, or using them only for data imputation or only for initial estimations (not both, and in their absence replaced by mean data or SIMPLISMA [25] profiles, respectively), the MCR models faced convergence issues. The results obtained were not satisfactory, even when trilinearity constraints were applied. These details have not been shown here and are part of future work, but they seem reasonable considering the complexity of the structure to be resolved by the bilinear model. Finally, it is worth noting that even when the experimental data are exactly the same, the structure used for processing them (2-way arrays, 3-way arrays, pseudoArrays, etc.) can be crucial in achieving good results, beyond differences in algorithm implementation.

# 4. Conclusions

Two datasets of fluorescence signals were studied, obtained through sequential emission scans at different excitation wavelengths, with simultaneous concentration variations. As this type of scanning requires time, the information modes can change simultaneously over time. This can generate dependencies among modes, which are sometimes incompatible with conventional multilinear models and/or certain data structures.

The results from models with variable concentrations, as well as some findings related to the photodegradation of the fluorophores themselves, suggest that acquiring conventional EEMs may present challenges. If concentrations are not kept relatively constant from the beginning to the end of an EEM, such fluorescence data can be forced into a 2-way array but will not exhibit conventional bilinearity. However, when partially grouped into short time intervals and localized within them, the same data can even be considered trilinear.

Some important properties of the data may only be useful at certain time scales, and time measurements can help with this. Models based on psCubes (or on matrices derived for MCR) with measured and constrained times proved to be appropriate for better representing the data. The data structure and its underlying model, as well as time-based smoothing using the specific time points of each sample, are key to resolving datasets such as those studied here, especially in the presence of a high percentage of missing data.

The complexity associated with resolving data structures like those presented requires good initial estimates and imputations to converge to realistic solutions. Previously resolved profiles and strategies involving profiles with gradually increasing time resolution can be employed to assist MCR.

# CRediT authorship contribution statement

**Gabriel Siano**: Conceptualization, Methodology, Software, Formal Analysis, Investigation, Writing – original draft, Writing – review & editing, Visualization

**Sofía Mora**: Methodology, Investigation, Data curation, Writing – original draft

**Agustina Schenone**: Visualization, Resources, Writing – review & editing

**Leonardo Giovanini**: Formal Analysis, Resources, Writing – review & editing

# Acknowledgments

We are grateful to Universidad Nacional del Litoral for its financial support (CAID 85520240100089LI). G.S. extends gratitude to Patricia Cortés for some discussions related to the results of these experiences.

# Supplementary Information

**Table of contents**

| | |
|---|---|
| Explained Variance (%) | 94.0307 |
| **Analyte (Pyr) Predictions (ppm)** | Predicted / Nominal |
| Validation Sample 1 | 39.56 / 59.92 |
| Validation Sample 2 | 47.70 / 74.98 |
| Validation Sample 3 | 72.56 / 89.88 |
| Validation REP% | 29.8774 |
| Validation RMSE (mg L-1) | 22.0510 |
| Validation Mean Recovery % | 70.1226 |
| **Criterion of similarity (0 to1)** | |
| Excitation Pyr | 0.9823 |
| Excitation 4A | 0.8755 |
| Excitation Trp | 0.9844 |
| Excitation Tyr | 0.9394 |
| Emission Pyr | 0.9998 |
| Emission 4A | 0.5997 |
| Emission Trp | 0.9894 |
| Emission Tyr | 0.8791 |
| mean spEX | 0.9454 |
| mean spEM | 0.8670 |

**Table S1**: **Conventional MCR model**. Qualitative and quantitative descriptors for the LC-EEM system.

Pyr: Pyridoxine; 4A: 4-aminophenol; Trp: Tryptophan; Tyr: Tyrosine

spEM: emission spectrum; spEX: excitation spectrum

MCR models: Assuming independence between excitation and elution time modes
Data classification: type 1 non-quadrilinear third-order data
Dependence between modes: samples/elution time
Common Mode: concatenation of excitation and emission modes
Super-augmented Mode: samples and elution time modes
Example in LC-EEM literature: Online Third-Order Liquid Chromatographic Data with Native and Photoinduced Fluorescence Detection for the Quantitation of Organic Pollutants in Environmental Water.
Model constraints (the same as in the LC-EEM literature example): Non-negativity in all modes, area correlation and correspondence between components and samples. The unimodality constraint was applied to the elution profile.
- In all cases, the MVC3 Toolbox was utilized.
- For all models, initial profiles were obtained by estimating the purest variables in the respective common mode.

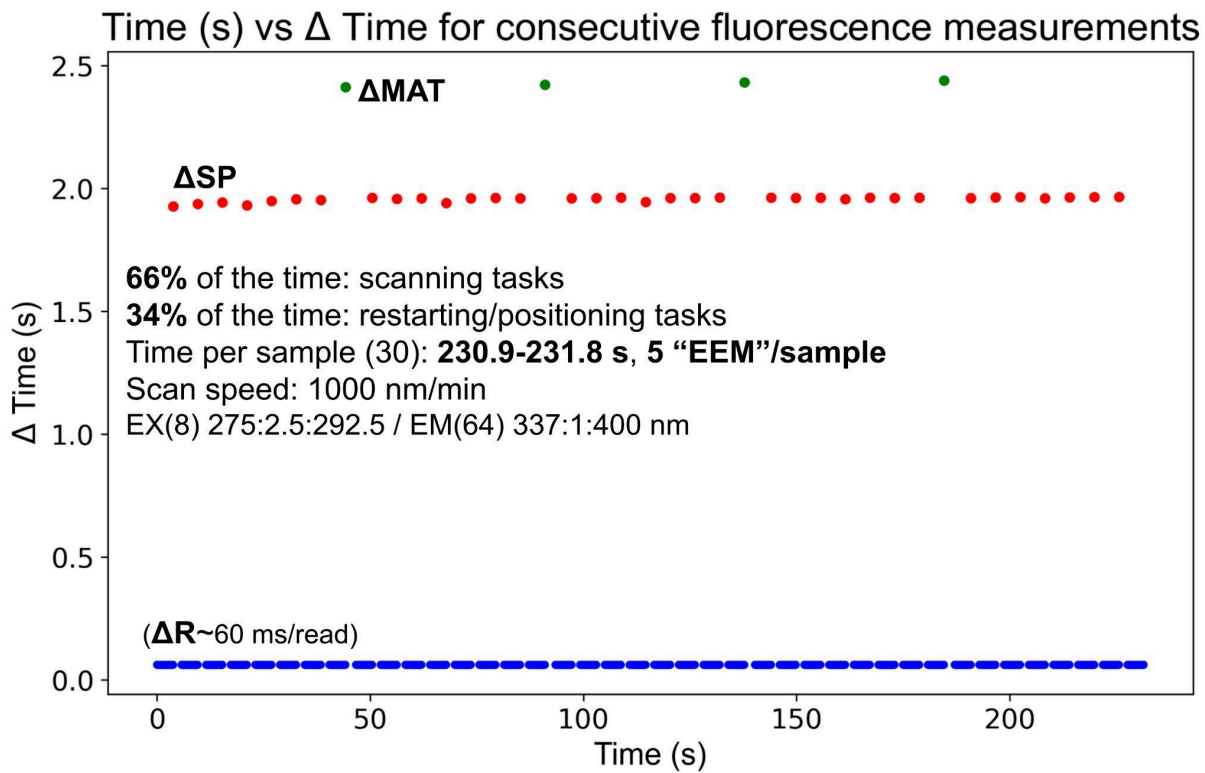

**Figure S1**: Time measurements for a representative (calibration) sample of the Kin-EEM system

ΔR (blue): time required for each individual reading within an emission scan; ΔSP (red): time difference between the end of a scan and the start of the next; ΔMAT (green): time difference between the end of an EEM and the start of the next.

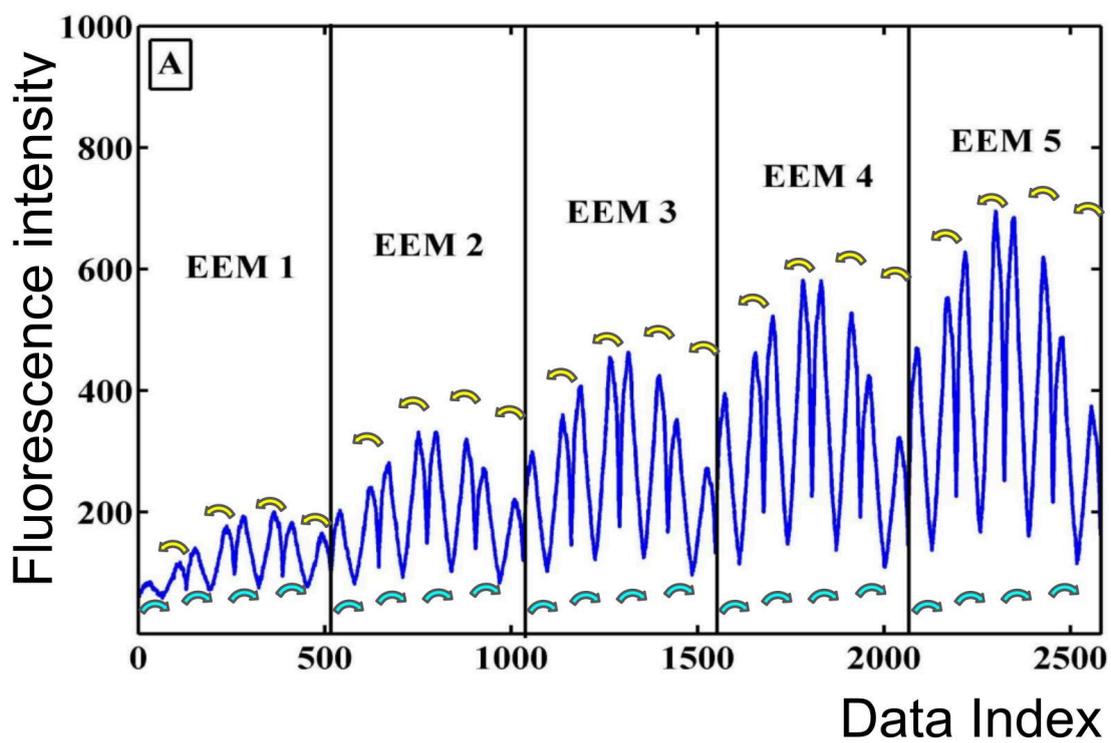

**Figure S2**: Consecutive emission scans for a pure Diclofenac sample.
Cyan arrows: emission scans oriented towards increasing wavelengths; Yellow arrows: emission scans oriented towards decreasing wavelengths

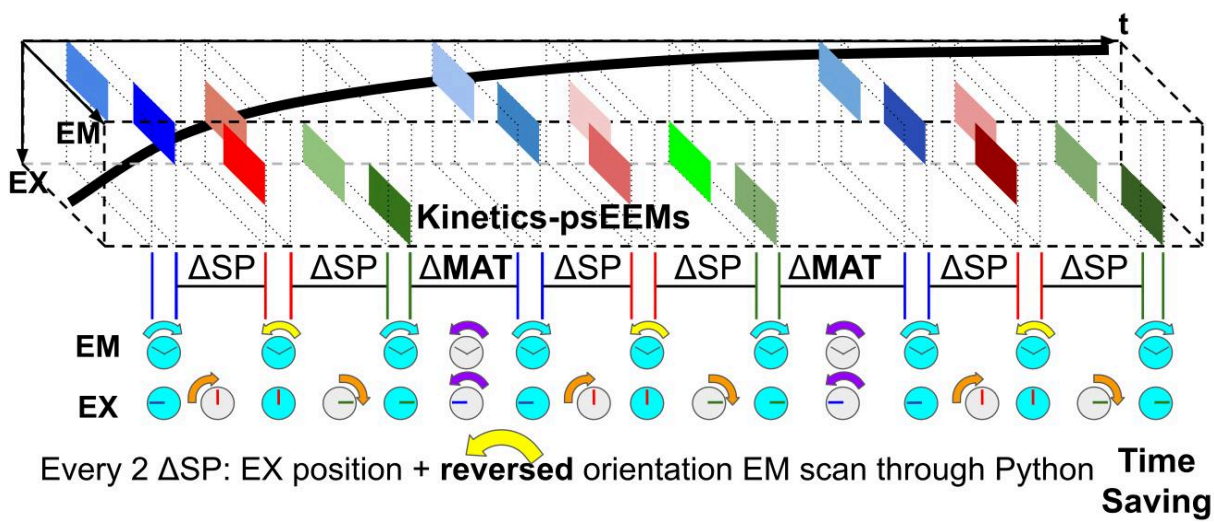

**Figure S3**: Schematic of monitoring the kinetics of photoderivative formation from Diclofenac using psCubes with EM scans.
**Top**: Three consecutive EEMs represented as three consecutive psCubes, each composed of six psEEMs. Each psEEM, outlined with dotted lines, contains only half of an spEM (colored rectangles, either the initial or final half of a scan) for a single EX variable (blue, red, or green tones), while the remaining data are treated as missing. Behind the augmented psCube, in bold black lines, is a hypothetical kinetic profile.
**Bottom**: Time differences between spectra (ΔSP) and between matrices (ΔMAT), along with schematic representations of restarting, positioning, and scanning processes for hypothetical EX and EM monochromators.

For the real case, the augmented psCube for each sample was constructed from 5 EEMs. Each EEM had dimensions of 8 in EX and 64 in EM, and 1/4 of an spEM was used for each psEEM (in the schematic, this would correspond to 1/2). As a result, the augmented psCube for each sample consisted of 160 time points, 8 EX, and 64 EM. It is important to note that the red-toned schematics are inverted in time relative to the blue and green tones. Before processing the data, the orientation of these spectra must be reversed. This is feasible because it is assumed that fluorophore concentrations did not vary, making the scan orientation irrelevant.

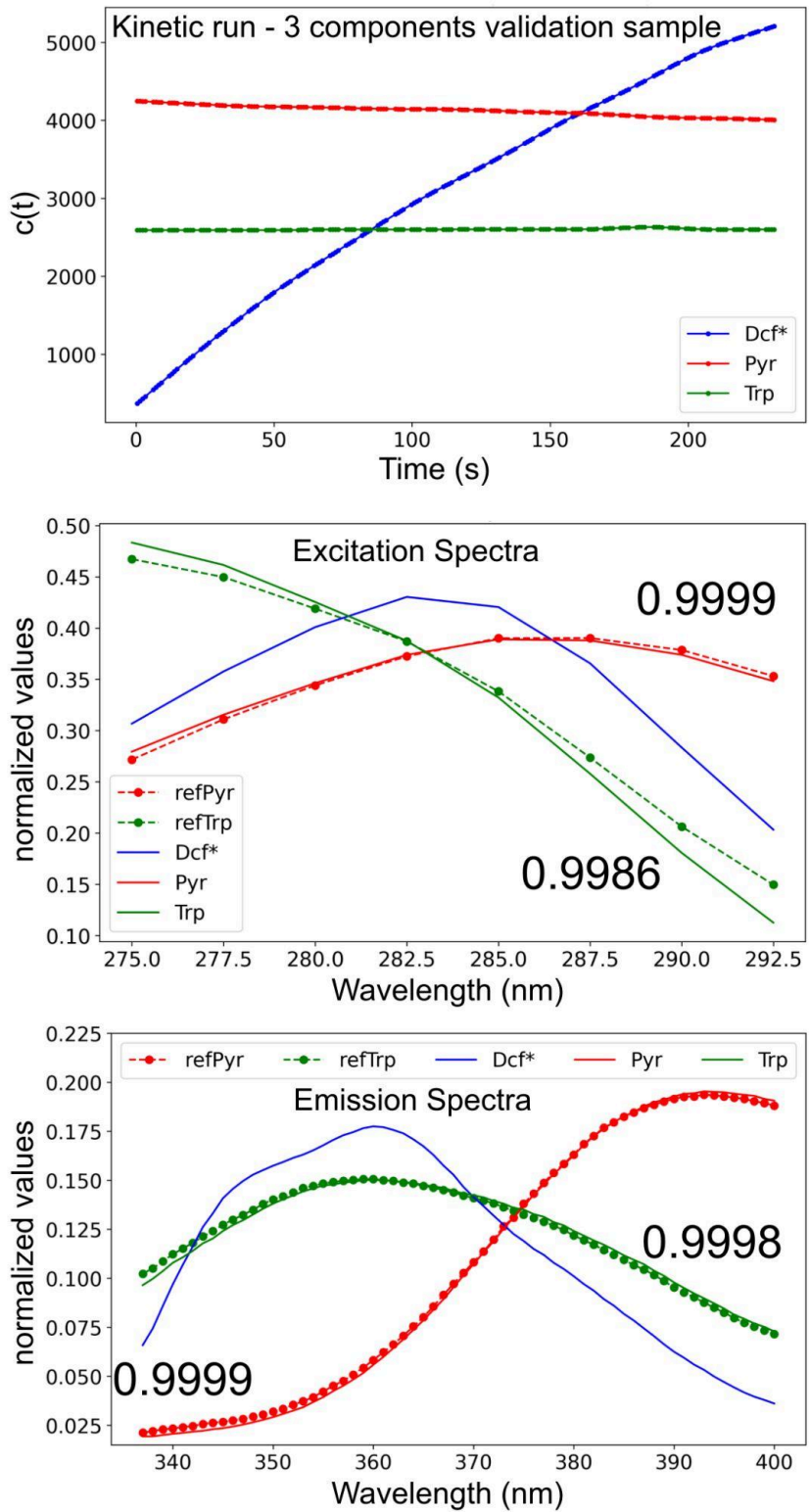

**Figure S4**: KinEEM, time assisted MCR model results for a validation sample with 3 fluorescent components.
**Top**: kinetic profiles resolution; **Middle and bottom**: spectral profiles resolutions (normalized values). Dcf*: photo-product of DCF; Pyr: Pyridoxine; Trp: Tryptophan; the prefix "ref" means reference (similarity coefficients are indicated for some spectra against their references)

| Calibration | |
|---|---|
| Equation | 92.8·[ppm Pyr] + 468 |
| $R^2$ | 0.996 |
| Working range | 0-100 ppm |
| EJCR test | OK |
| **Analytical Figures of Merit** (based on pseudo-univariate graph) | |
| Sensitivity | 92.8 ppm$^{-1}$ |
| Analytical sensitivity | 0.46 ppm$^{-1}$ |
| LOD | 5 ppm |
| LOQ | 15 ppm |

**Table S2**: LC-EEM calibration model details

| Calibration | |
|---|---|
| Equation | 9292·[ppm DCF] + 2539 |
| $R^2$ | 0.990 |
| Working range | 0-8 ppm |
| EJCR test | OK |
| **Analytical Figures of Merit** (based on pseudo-univariate graph) | |
| Sensitivity | 9292 ppm$^{-1}$ |
| Analytical sensitivity | 3.50 ppm$^{-1}$ |
| LOD | 0.73 ppm |
| LOQ | 2.2 ppm |

**Table S3**: Kin-EEM calibration model details

| Sample Number | [Diclofenac]$_{ppm}$ | [Pyridoxine]$_{ppm}$ | [Tryptophan]$_{ppm}$ |
|---|---|---|---|
| 1-2-3 | 2 | 0 | 0 |
| 4-5-6 | 4 | 0 | 0 |
| 7-8-9 | 6 | 0 | 0 |
| 10-11-12 | 8 | 0 | 0 |
| 13-14-15 | 4 | 0 | 0 |
| 16-17-18 | 4 | 0.3 | 0 |
| 19-20-21 | 0 | 0.3 | 0.3 |
| 22-23-24 | 5 | 0.3 | 0.3 |
| 25-26-27 | 0 | 0 | 0.3 |
| 28-29-30 | 0 | 0.3 | 0 |

**Table S4**: KinEEM, Composition of calibration and validation samples

The concentrations of the analyte and interferents were selected to ensure significant fluorescence units without saturating the detector when measuring mixtures. Additionally, higher concentrations that could lead to extra phenomena, such as inner filter effects or quenching, were avoided. Furthermore, the fluorescence of the interferents was already known.

Samples 1 to 12 and 22 to 30 were prepared using pure standards of diclofenac, pyridoxine, and tryptophan.

Samples 13 to 15 were prepared using the injectable OXA® B12 I.M., which contains diclofenac, betamethasone, and hydroxocobalamin (vitamin B12). Under the working conditions, only diclofenac exhibits fluorescence.

Samples 16 to 18 were prepared using the injectable OXA® B12 I.M., with the addition of pyridoxine (vitamin B6). Pyridoxine was included to simulate its combination with diclofenac, as found in medications such as Dioxaflex Complex (Laboratorios Bagó®, Buenos Aires, Argentina).